\documentclass[epj]{svjour}

\usepackage{graphics}
\usepackage{epsfig}
\usepackage{graphicx}
\usepackage{epsfig}
\usepackage{amssymb}
\usepackage{amsmath}
\newcommand{\bc}{\begin{center}}
\newcommand{\ec}{\end{center}}
\newcommand{\be}{\begin{equation}}
\newcommand{\ee}{\end{equation}}

\newcommand{\pic}[1]{\includegraphics[width=90mm]{#1}}
\newcommand{\picc}[1]{\includegraphics[width=90mm,height=55mm]{#1}}
\newcommand{\tr}{\hbox{tr}}

\begin{document}
\title{Glueballs and the superfluid phase of Two-Color QCD}
\author{M.P. Lombardo \inst{1}\and
M.L. Paciello \inst{2} \and
S. Petrarca \inst{2,3} \and
B. Taglienti \inst{2} } 

%
%
\institute{INFN, Laboratori Nazionali di Frascati, via E. Fermi 40,
I00044 Frascati, Italy \and
INFN, Sezione di  Roma,
P.le A. Moro 2, I-00185 Roma, Italy \and
Dipartimento di Fisica {\it{Sapienza}} Universit\`a di Roma, \\P.le A. Moro 2, I-00185 Roma, Italy}
%
%
\abstract{
We present the first results on scalar glueballs in cold,  dense
matter using lattice simulations of two color QCD.
The simulations are carried out on a $6^3 \times 12$ lattice and use a
standard hybrid molecular dynamics algorithm for staggered 
fermions for two values of quark mass. The  
glueball correlators are evaluated via a multi-step smearing procedure.
The amplitude of the glueball correlator peaks in correspondence with
the zero temperature chiral transition, $\mu_c = m_\pi/2$, and
the propagators change in a significant way in the superfluid
phase, while the Polyakov loop is mearly insensitive to the
transition.  Standard analysis suggest that 
lowest mass in the $0^{++}$ gluonic channel decreases in the 
superfluid phase, but these observations need to be confirmed
on larger and more elongated lattices  These results indicate that
a nonzero density induces  nontrivial modifications of the gluonic medium.
} 
\maketitle
\section{Introduction}

The study of the phases of QCD has applications to a wide array of subjects,
ranging from heavy ion collision from FAIR to RHIC to LHC energies, 
the physics of compact stars, QCD phase transitions in the early universe,
and more theoretical scenarios in which strongly coupled gauge theories
might serve as a laboratory for physics beyond the standard model~\cite{Alford:2007xm}.

Lattice field theory provides a solid computational framework for these
studies, in  all those cases in which the determinant is 
positive and importance sampling can be applied.
Unfortunately, at high baryon density the weight of the functional
integral is complex, which forbids the usage of simple importance
sampling in the computation. 

At high temperature strong fluctuations of the baryon number makes it possible
to study the physics of nonzero baryon density, and in this way we can explore
the portion of the phase diagram which is analytically connected with the
zero density quark gluon plasma region. Unfortunately, this
excludes many different other phases of QCD, which appear at low temperatures
and higher densities, see e.g.~\cite{Stephanov:2007fk} for an overview.

Two color QCD is an important exception:  the color SU(2) group 
in the fundamental representation is pseudoreal and in this
case the determinant is real even when $\mu \ne 0$ and
standard hybrid Monte Carlo simulations
are possible.  So
lattice calculations can confront and extend fermionic model studies~\cite{Sinclair:2003rm}.

However, very little happened in the field after the pioneering studies
of the 80's ~\cite {Kogut:1983ia,Nakamura:1984uz,Dagotto:1986gw}.
 The subject was revived after the work of the late 90's~\cite{Alford:1997zt,Rapp:1997zu}:
these authors  noted that phenomena associated by
diquark condensations at high density might well be much more important
than previously though, if one goes beyond one gluon exchange when
calculating the relevant four fermion couplings.
Two color QCD was then reconsidered as a laboratory for a toy study
of diquark condensation mechanisms~\cite{Hands:1999md}.
Symmetries and spectrum were analysed
in a quantitative way, and lattice studies of diquark condensation were
carried out~\cite{Kogut:2001na,Aloisio:2000rb}. 
Further studies include RMT analysis \cite{Akemann:2004dr,Splittorff:2007ya}
and studies of the Dirac spectrum \cite{Bittner:2000rf}.
Moreover, the fact that $\mu_c(T=0) = m_\pi/2$
makes the model  amenable to a $\chi$PT analysis~\cite{Kogut:2000ek}.
In turn, chiral perturbation theory motivated studies produced
results for the EoS and Gluon Condensate~\cite{Zhitnitsky:2007uk}.
The model (with adjoint fermions) has been also used as a laboratory for the study of
new universality classes ~\cite{Sannino:2004ix}. 
The phase diagram has been computed in the chiral limit by use a new
cluster algorithm~\cite{Chandrasekharan:2006tz}.
Model studies of the vector sector~\cite{Sannino:2002wp}
prompted studies of  vector spectroscopy on the 
lattice~\cite{Alles:2002st,Muroya:2002ry,Hands:2004ry,Hands:2007uc}.

In short summary, model analysis, chiral perturbation theory, and the
feasibility of lattice calculations conspired to make the fermionic
sector of two-color QCD reasonably well understood, with numerical results
in quantitative agreement with analytic calculations.

In comparison with the fermionic sector, the
 gluodynamics of two-color QCD is much less well known.
Studies have been carried over at high temperature, where
the finite $\mu$ transition is similar to the one at finite T
~\cite{Alles:2006ea,Conradi:2007kr}, while 
at low temperature the behaviour of the Polyakov loop and of the
gluon propagator
indicates that the hadronic/superfluid transition and the
deconfining transition are well separated phenomena~\cite{Hands:2006ve}.

In this study we present a quantitative analysis of the scalar 
plaquette-plaquette correlator, with the main goal of making progress
on our understanding of the phase diagram of QC$_2$D at low temperature,
and to get further insight into the nature of the gauge field dynamics 
at high baryon density, in a region which is distinctly different from the
ordinary, high temperature, high baryon density phase.
Glueballs have been considered as probes of the thermal medium 
in~\cite{ishii1,ishii2}.

Previous work on the subject has discussed at length the similarities
and differences between the phase diagrams of two and three colors QCD.
In short, while the properties of the matter above $T_c$ are substantially 
independent on the number of colors 
the nature of the low temperature, high density phase seems to 
depend strongly on the number of colors.
At the same time, though, many of the differences appear in the fermionic
sector, where different couplings produce different pattern of the
symmetries realisation. The gluonic sector is not directly related with
these different symmetries, and might well exibit characteristics
independent on  the number of colors. This is one of the motivation
for a lattice study.
On a more general grounds, two color QCD at high density affords an
example, as we will see, of a chiral transition clearly separated
from the deconfining one. 

This paper is organized as follows: in the next Section  we review a few 
basic facts about two color QCD and its phase diagram; Section~\ref{sec:simulation}
is devoted to the details of the simulations, while Section~\ref{sec:measurements} deals
with observables and measurement strategies. Smearing~\cite{smearing} 
is introduced
and reviewed there. Section~\ref{sec:results} contains the results: 
we discuss the behaviour of the scalar gluonic correlators  constructed
by the smearing technique 
in order to excite glueballs and study 
their behavior as a function of baryon density. Results for the
Polyakov loop are presented 
as well. In Section~\ref{sec:summary} we summarise 
and discuss our findings, and outline future research. 

A preliminary account of some of the results have 
appeared in~\cite{lattice1,lattice2}.

\section{Action, symmetries, phase diagram}\label{sec:phaseDiag}

Let us briefly remind here a few  features of the thermodynamics
of the two color model on which we will rely in the following. 
The above mentioned reviews
and literature can of course be consulted for more details.

The lattice version of the Action reads~\cite{Hands:1999md}:
\begin{equation}
\begin{split}
S_{kin}= \sum_{x,\,\nu=1,3}{\frac{\eta_\nu(x)}{2}} 
[&\bar\chi(x)U_\nu(x)\chi(x+\hat\nu)- \nonumber \\
 &\bar\chi(x)U_\nu^\dagger(x-\hat\nu)\chi(x-\hat\nu)]+  \nonumber \\
\sum_{x}{\frac{\eta_t(x)}{2}}
[&\bar\chi(x)e^{\mu}U_{t}(x)\chi(x+\hat{t})- \nonumber \\
 &\bar\chi(x)e^{-\mu}U_{t}^\dagger
(x-\hat{t})
\chi(x-\hat{t})] \nonumber
\label{eq:sfund}
\end{split}
\end{equation}
where a chemical potential per quark was introduced in the usual
way\cite{Kogut:1983ia,Hasenfratz:1983ba}.

At $\mu=0$ the Action enjoys the lattice equivalent of Pauli-G\"ursey symmetry:
\begin{equation}
\tau_2U_\mu\tau_2=U_\mu^*
\end{equation}
where $\tau_2$ is Pauli matrix .

The Action can be recast
\begin{equation}
\begin{split}
S_{kin}=&\sum_{x\;even,\,\nu=1,3}{\frac{\eta_\nu(x)}{2}}
[\bar X_e(x)U_\nu(x)X_o(x+\hat\nu)-  \nonumber \\
 &\bar X_e(x)U_\nu^\dagger(x-\hat\nu) X_o(x-\hat\nu)] + \nonumber \\
 &\sum_{x\;even}{\frac{\eta_{t}(x)}{2}}
[\bar X_e(x)(
\begin{matrix} e^{\mu} &0\cr 0& e^{-\mu} 
\end{matrix}
) U_{t}(x)X_o(x+\hat{t})
-  \nonumber \\
&\bar  X_e(x)(
\begin{matrix}
e^{-\mu}&0\cr0&e^{\mu}
\end{matrix}
)U_{t}^\dagger(x-\hat{t})
X_o(x-\hat{t})]
\label{eq:sfundX}
\end{split}
\end{equation}
with the definitions
\begin{equation}
\bar X_e=(\bar\chi_e,-\chi_e^{tr}\tau_2)\;\;:\;\;
X_o=(
\begin{matrix}
\chi_o\cr -\tau_2\bar\chi_o^{tr}\cr
\end{matrix}
).
\label{eq:Xfields}
\end{equation}
which make even more transparent the equivalence of quarks and antiquarks
in the model.

At $\mu=0$ The (Pauli-G\"ursey) symmetry
implies the degeneracy of mesons and baryons (diquarks) with opposite parities:
\begin{equation}
m_\pi = m_{qq}
\end{equation}
In turn, this produces  only one discernible condensate, 
$<\bar \psi \psi>^2 + <\psi \psi>^2$, whose direction is chosen by the mass
 term. 

Consider now the thermodynamics of the model.

At zero density, high temperature there is a transition separating the
ordinary hadronic phase from a quark gluon plasma. 
The characteristics of the high temperature and low temperature
phases seem to be fairly independent on the number of colors, with $T_c$ of
the order of a typical mass scale (of course the precise
nature of the transition is not). This remains true for a nonzero
chemical potential, till the temperature if ``high enough'', most likely
till the transition is between a hadronic phase and a quark gluon plasma
phase.

At zero temperature standard arguments indicate that the there is no Fermi sea
until the quark chemical potential exceeds the mass of the lowest state
which carries baryon number, i.e half the mass of the baryon which is 
degenerate with the pion: $\mu_c = m_\pi/2$.  
For $\mu > \mu_c$ 
the attractions in the diquark channel leads to the formation of a diquark
condensate $<\psi \psi >$. For two color QCD $<\psi \psi >$ is color neutral, 
like the chiral condensate itself: when $\mu$ exceeds $\mu_c$ ,
$<\bar \psi \psi > $ rotates to $<\psi \psi>$, and numerical simulations
nicely follow the prediction of chiral perturbation theory for the chiral
condensate, mass spectrum and thermodynamics.

\begin{figure}
\pic{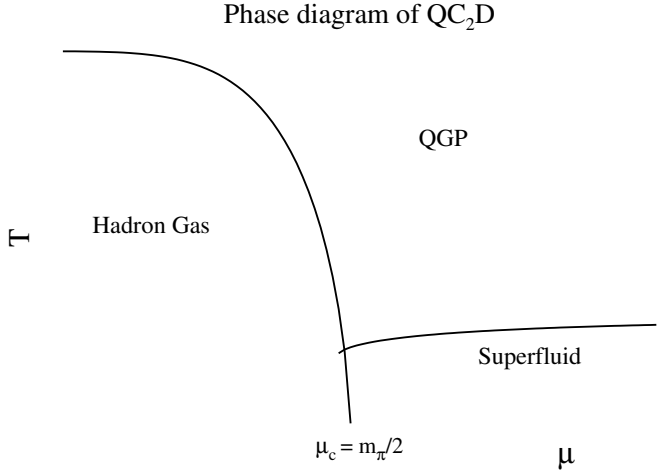}
\caption{Sketchy view of the phase diagram of two color QCD: we recognize
(at least) three phases. A Hadronic phase, a BEC phase, and a Quark Gluon 
Plasma phase. The phases are characterised by the values of the 
$<\bar \psi \psi>$ and $< \psi \psi >$ condensates, the Polyakov loop and
the number density. (In this plot we do not distinguish between transition
and crossover lines) }
\label{fig:phase}
\end{figure}

\section{The simulation}\label{sec:simulation}
\label{sim}

The simulations were performed using a standard
hybrid molecular dynamics algorithm for staggered fermions using the
same code as of ref. 
~\cite{Hands:1999md}, which corresponds to $N_f = 8$ continuum flavors.

Details on relevant parameters are summarised in table~\ref{tab:param}.
Smaller $dt$ were needed in the supercritical region and trajectory
lengths ranged in most cases from half to one. 

\begin{table}[h]
\caption{Simulation parameters} 
\label{tab:param}
\vskip 5mm
\begin{tabular}{|l|l|}
\hline
                        & \\
lattice size            &  $6^3 \times 12$\\
$\beta$                 &  $1.3$\\
configurations          &  $40000$\\
sweeps  &  $10$\\
dt      & 0.05 - 0.02 \\
thermalization          &  $1000$ configurations\\
quark mass      $m_q$        &  $.05,.07$\\
chemical potential   $\mu$   &  $.0,.2,.25,.30,.32,.35,.4,.6,.9$\\
smearing weights $\omega$  &  $.025,.05,.07,.1,.2,.3,.4$\\
smearing steps          &  $1,2,3,4$\\
                        & \\
\hline
\end{tabular}
\end{table}

Values $\mu=0.25, 0.30, 0.32, 0.35, 0.90$ were simulated only at $m_q=0.07$,
with a slightly reduced statistics of $30000$ steps. For the analysis
we used a bin size of 7000, for which the errors level off.

To set the scale, 
we have measured the fermionic spectrum at $\mu=0.0$.

\begin{figure}
\bc
\pic{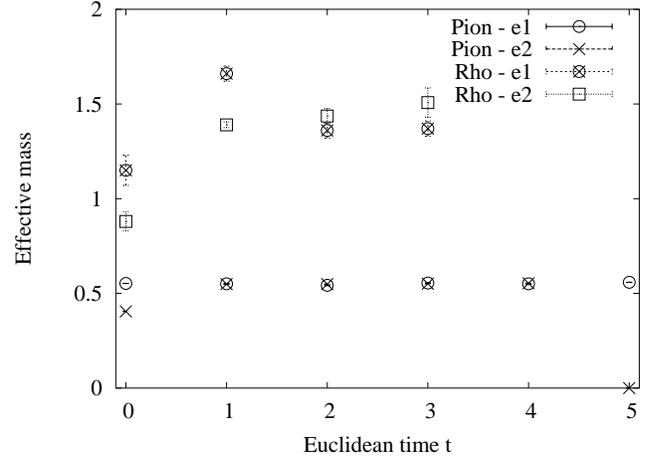}
\caption{Effective masses for the $\rho$ and the $\pi$ mesons 
in the normal
phase ($\mu=0.0$) and bare quark mass $m_q= 0.05$,
from the two different procedures labeled as e1 and e2, see text.}
\label{fig:smear}
\ec
\end{figure}

We have constructed
the zero momentum propagators $C(t)$ for the pion and the $\rho$ mesons
 in the standard way, 
 assumed the  form 
\begin{equation}
C(t) = A \cosh [m (t- N_t/2)] 
\label{eq:efe}
\end{equation}
which is valid for euclidean times $N_t$  larger than the reciprocal of
the excited masses,
and used two different definition of the (time dependent) 
effective masses $m_{eff}(\bar t)$, 
 one  exploiting  $C(t)/C(t+1)$
\begin{equation}
C(\bar t)/ C (\bar t   + 1) = \frac 
{\cosh [m_{eff}(\bar t) (t - N_t/2)]}  {\cosh [m_{eff}(\bar t) (t + 1 - N_t/2)]}  
\label{eq:eff}
\end{equation}
the other  $C(t)/C(t+2)$, according to
\begin{equation}
C(\bar t)/ C (\bar t   + 2) = \frac 
{\cosh [m_{eff}(\bar t) (t - N_t/2)]}  {\cosh [m_{eff}(\bar t) (t + 2 - N_t/2)]}  
\label{eq:ef2}
\end{equation}
We show 
 the results for the $\rho$ and the pion at $m_q = 0.05$,
in Fig.~\ref{fig:smear}, where  
we label the effective masses extracted by use of the two definitions
as e1 and e2 respectively. The two procedures give 
comparable results on a moderate range of times, indicating that the
oscillatory  
contribution for the opposite parity partner, if any, is very small. 

>From the plateaus in the plots 
we read off $m_\pi = 0.56 (2)$ and $m_\rho = 1.4(1)$. 
A similar analysis  at $m_q = 0.07$ gives $m_\rho = 1.5(2)$ 
and $m_\pi = 0.64$. In both cases
$m_\rho$ and $m_\pi$ are well resolved, meaning that  we are still 
within the range of chiral perturbation theory.

\section{Glueball measurements: operators and smearing}\label{sec:measurements}
\label{op}
The operators commonly used for measuring scalar gluonic correlators exciting 
glueball masses are Wilson loops. For simplicity we restricted ourselves to 
plaquette-like operators that can be built from four links. 

Simple glueball wave functions such as the plaquette have the 
disadvantage that they only have small overlaps with the lowest-lying 
glueball  states, and that these overlaps become rapidly smaller as the 
lattice spacing 
is decreased. 

Furthermore  the plaquette couples strongly to ultraviolet 
fluctuations, increasing the noise in the correlators;
to have reliable glueball correlation functions at different distances, 
it is mandatory to reduce these statistical fluctuations. 

Different smoothing procedures aimed at removing 
the unphysical short-distance fluctuations 
have been introduced; a short review can be found in~\cite{review}.
One of this is the smearing method, originally 
proposed for SU(3) in~\cite{smearing}. It  consists in the construction of 
correlation functions for operators which are 
a functional of the field smeared in space and not in time.  

For each 
link $U_\mu(x)$ of a configuration the product of the other three links 
defining a 
plaquette is considered, then these products are summed over the four 
choices of plaquettes orthogonal to the time axis; 
the resulting matrix is added
with a weight $w$ to the original link $U_\mu(x)$. This linear combination,  
projected back on the gauge group, is assumed as new link variable 
(the so called ``fat link''). The procedure can be iterated, say $N_s$
times.

This operation both enlarges the ``size'' of the ordinary plaquette operator 
and reduces the noise associated with the additional degree of freedom due 
to the gauge invariance.  Only spatial links participate on the averaging. 
Thus the transfer
 matrix for the smeared operators is unaffected and it
remains positive definite, if this was the case. 
The values of the smearing coefficient $w$ and the
number of iterations $N_s$  are  tuned in order 
to optimise the performance of the method \cite{alba,calibro,gupta}.

The glueball operators are defined by means of the plaquettes $P_{ij}(\vec{x},t)$ on $ij$ plane as in the following:
\begin{equation}
\phi ^{0^{++}} (t) = \tr \sum _{\vec{x} } \biggl[ 
P_{12}(\vec{x},t) \, + \, P_{23}(\vec{x},t) \, + \, 
P_{13}(\vec{x},t) \biggr] 
\label{eq:oper0}
\end{equation}
which transforms according to the $A_{1}^{++}$ (one-dimensional) 
 irreducible representations of the of the relevant cubic point 
group  (on a lattice the full rotational symmetry is broken down to only 
cubic symmetry), and
\begin{eqnarray}
\phi_a ^{2^{++}} (t) &=& \tr \sum _{\vec{x} } \biggl[ 
P_{12}(\vec{x},t) \, - \, P_{13}(\vec{x},t) \biggr]
\label{eq:oper2a}\\
 \phi_b ^{2^{++}} (t) &=& \tr \sum _{\vec{x} } \biggl[ 
P_{12}(\vec{x},t) \, + \, P_{23}(\vec{x},t) \, - \, 
2 P_{13}(\vec{x},t) \biggr] 
\label{eq:oper2b} 
\end{eqnarray}  
which transform both according to the $E^{++}$ 
(2-dimensional)  representation.
Note that all operators are projected onto zero momentum by 
averaging over all spatial sites.

 These operators are the lattice counterparts  of
the continuum $O(3) \otimes Z(2)$, $J^{PC}$ = $0^{++}$ and $2^{++}$ 
respectively. Hence,   we can  label
the corresponding excitations as $0^{++}$ and  $2^{++}$ states.  
Of course this 
correspondence is not one to one but infinite to one. What is usually 
measured is the lowest excitation in the corresponding representation of the 
cubic group, corresponding to the fundamental glueball state.

Unfortunately, but not unexpectedly, the $2^{++}$ results turned out
to be too noisy to provide any significant information. 
Because of this, we will base the following analysis on the 
$0^{++}$ results alone.

With dynamical quarks, as in our case, mixing is possible between
purely gluonic and fermionic operators. So strictly speaking we
should not talk about ``glueball masses'', but, rather, of the lowest
excitation in that particular chanell, be it fermionic, gluonic or a
mix of the two. This said, for the sake of brevity, in the following
we will sometimes speak of ``glueballs'', 
meaning what has just been explained.

Following the methods  previously depicted, a smeared operator is obtained 
replacing the ordinary link variables $U(x)$ in the plaquette 
operator $P_{ij}(\vec{x},t)$  by the fat link $U'(x)$. 
Higher levels of smearing are obtained by varying the weight $w$ and by 
iterating the procedure $N_s$ times.

Glueball masses are calculated from the  Euclidean time 
behaviour of their 
correlation functions:

\begin{equation}
\begin{split}
 C^{0^{++}} (t) = \frac{1}{N_t} \sum_{\tau =0}^{N_t - 1}
 (& <\phi ^{0^{++}} (\tau + t) \phi ^{0^{++}} (\tau) >
 - \\
&<\phi ^{0^{++}} (\tau)>^2 )
\label{eq:c0d}
\end{split}
\end{equation}
where we exploited the cluster property
\begin{equation}
\lim_{t \to \infty} 
\phi ^{0^{++}} (\tau + t) \phi ^{0^{++}} (\tau) = \phi ^{0^{++}}(\tau) ^2
\end{equation}
 as well as
the translational invariance of $\phi ^{0^{++}}(\tau)$.
In the following whenever no confusion could arise we will refer
to the propagator eq.\ref{eq:c0d} simply as $C(t)$.

We have analysed the glueball correlations for different smearing parameters.
One observation which allows a more compact analysis, as well as
offers more control on the results, was proposed in 
ref.~\cite{calibro}. There, it has been shown that 
the two-dimensional parameters space of the number of 
sweeps $N_s$ and the smearing weight $w$  may be reduced  to a single 
dimension  via the parameter $T_s=N_s\times w$, in a certain range
$0  \le w \le w^{max}, 0 \le N_s \le N_s^{max}$, where $w^{max} $ and
$N_s^{max}$ depend, of course, on the specific application under consideration.

\begin{figure}
\pic{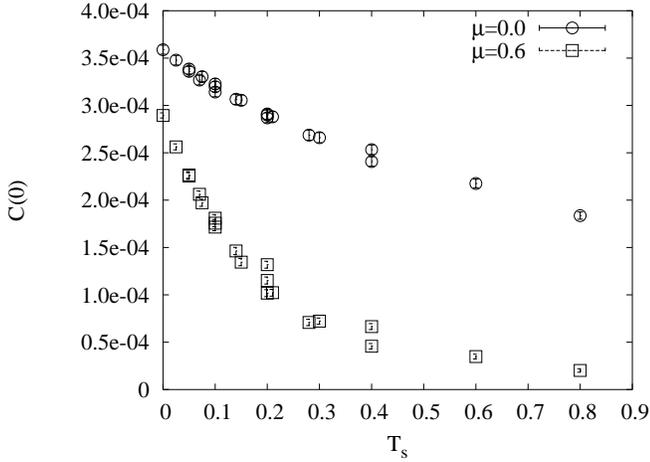}
\caption{Amplitude of the $0^{++}$ correlator for smearing weight $w$
and smearing steps $N_s$ as a function of the smearing parameter 
$T_s = N_s w$,  in the normal ($\mu=0.0$)
and superfluid phase ($\mu = 0.6)$, for $m_q = 0.05$.  
Note the universal behaviour
as a function of $T_s$ in the two phases.  Note also that smearing
affects the results at $\mu=0.6$ more than those at $\mu= 0.0$}
\label{fig:C0mu6mu0}
\end{figure}

In our case,  we have observed 
 that the unidimensional parametrisation remains true till 
$w \le 0.3$, irrespective of the number of smearing steps, at least
within the allowed range of smearing steps $N_s \le S/2$, where $S$
is the spatial size of the lattice. 
Fig.~\ref{fig:C0mu6mu0} demonstrates this property in the normal 
($\mu=0.0$) and in the superfluid phase 
($\mu=0.6)$ for the amplitude $C(0)$ of the correlators
built for our smallest quark mass $m_q = 0.05$. We have
observed that this behaviour holds true also for other observables,
and  mass $m_q = 0.07$.

To be on the safe side, we will base our 
discussions on results within the range of validity of the unidimensional
parametrisation: in conclusion, $w \le 0.3, N_s \le 3$.

The same Fig.~\ref{fig:C0mu6mu0} allows a comparison of the effects of
smearing in the two phases: we note that 
the smeared results drop much faster as function of $T_s$ 
after the transition ($\mu = 0.6$) than in the normal phase
($\mu = 0.0$). We will
come back to this point in the next section.

\section{Results}\label{sec:results}
We devote this Section to the presentation of the results
for the glueballs and the Polyakov loop.

\begin{figure}
\picc {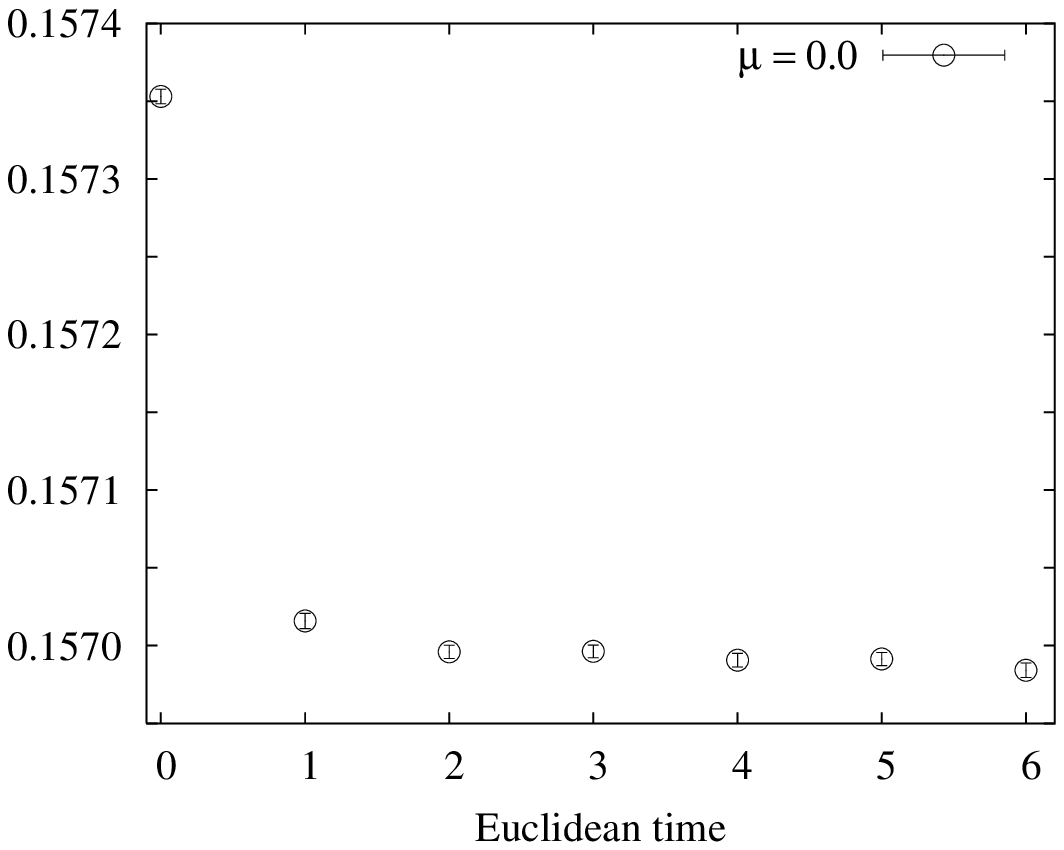}
\picc{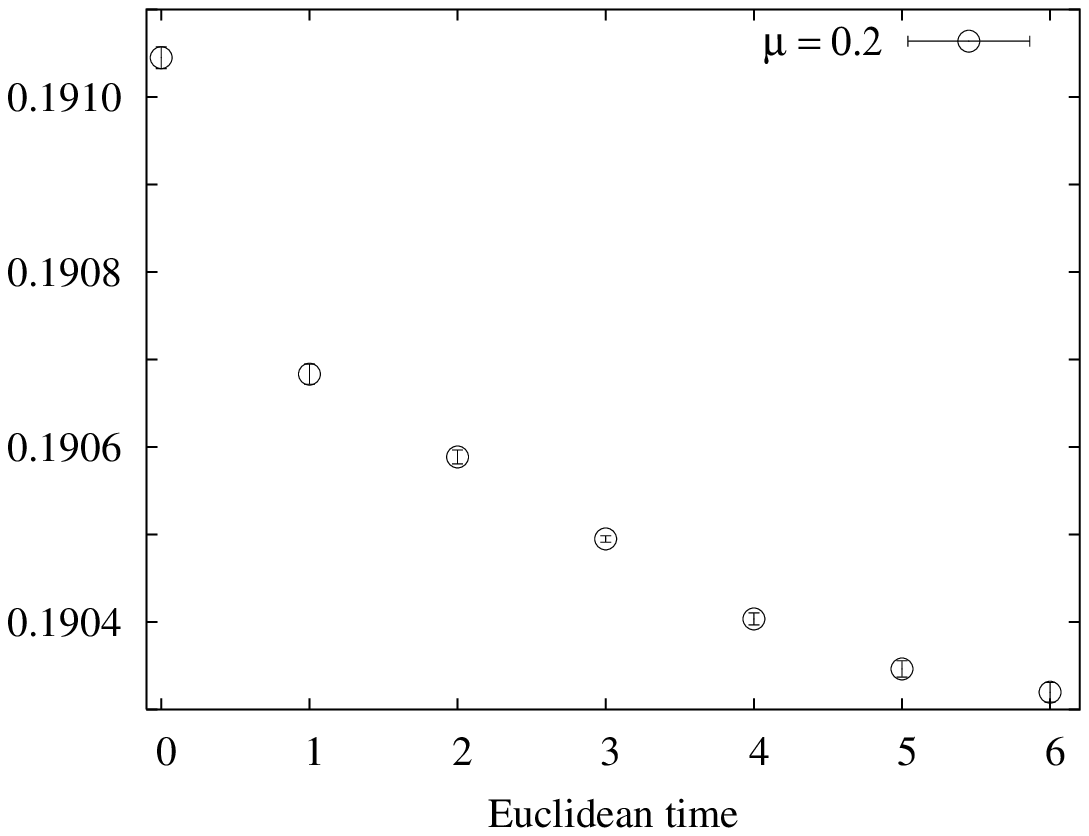}

\picc{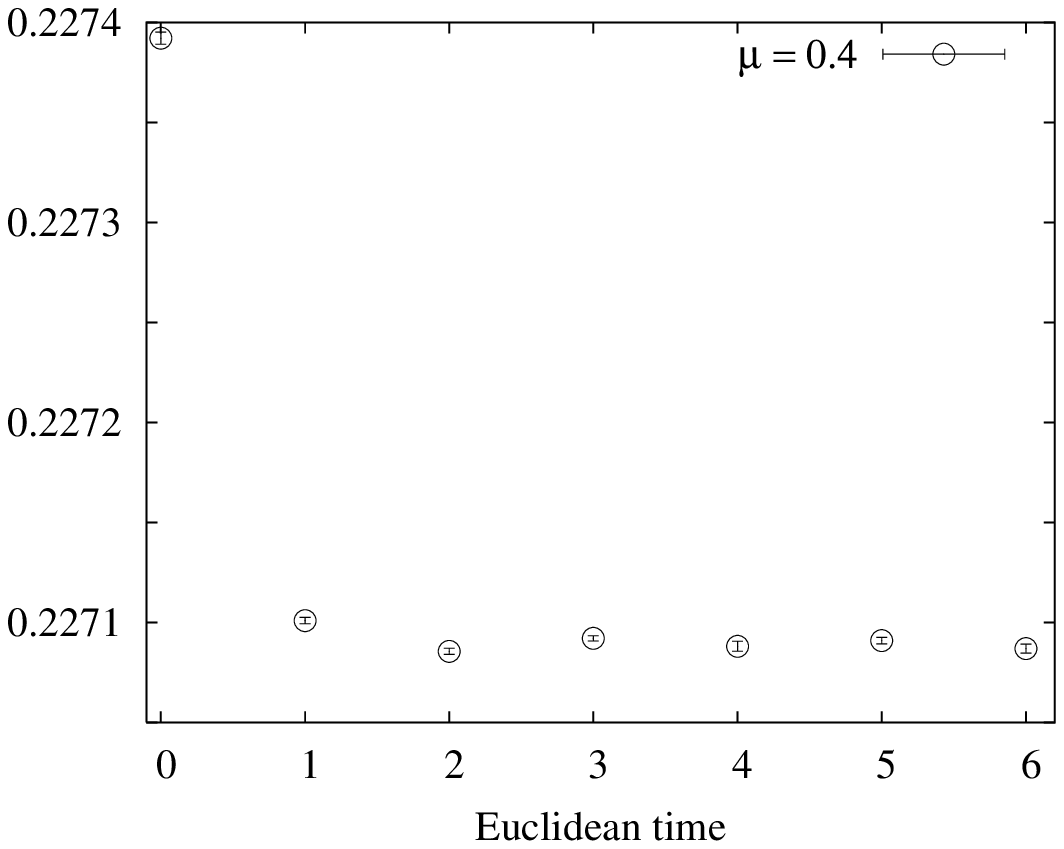} 
\picc{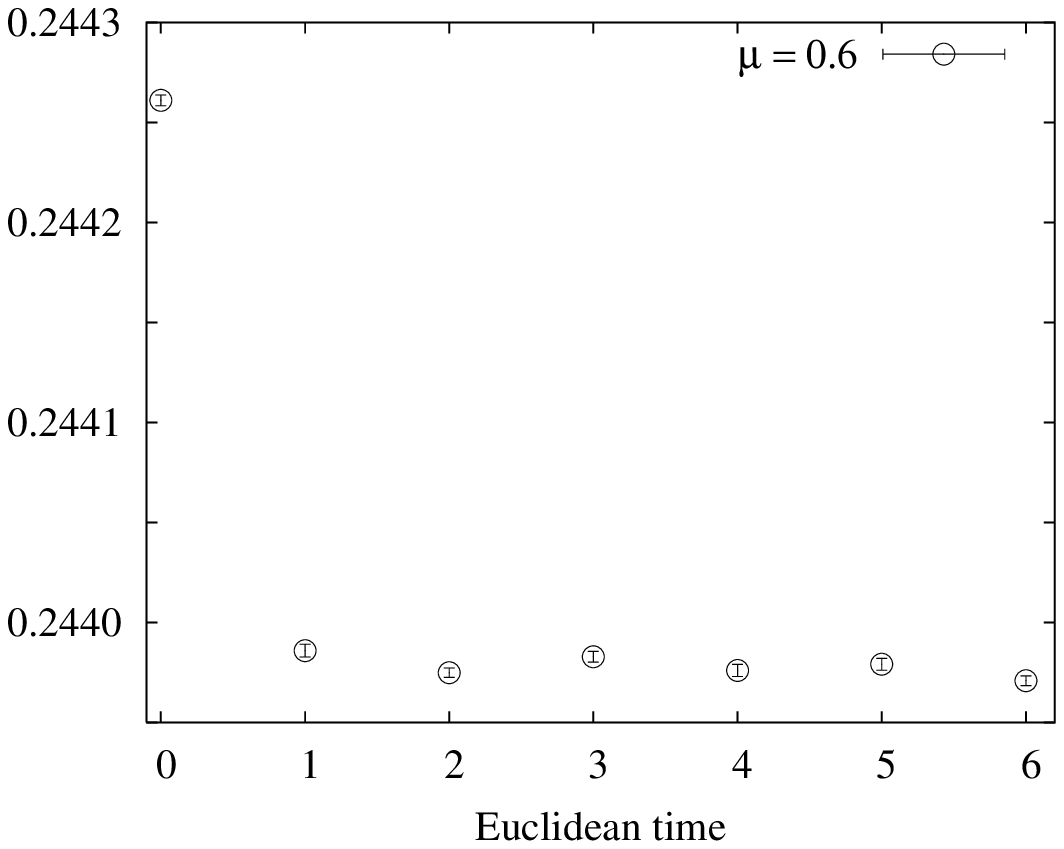}
\caption{Plaquette correlators (eq.\ref{eq:cc}) 
at mass $m_q = 0.05$ for a few $\mu$ values. $\mu=0.2$ is close to the critical
point, the two larger $\mu$ are in the superfluid phase.}
\label{fig:collection}
\end{figure}

The raw results for the plaquette correlators
\begin{equation}
 P (t) = \frac{1}{N_t} \sum_{\tau =0}^{N_t - 1}
 <\phi ^{0^{++}} (\tau + t) \phi ^{0^{++}} (\tau) >
\label{eq:cc}
\end{equation}
for
our smaller mass $m_q = 0.05$ are presented in Fig.~\ref{fig:collection}.
 The results in the two phases, the normal phase
at $\mu = 0.0$ and the superfluid phase, $\mu = 0.4, 0.6$,
show the expected long distance plateaux.

This suggests to base our analysis on the  propagators defined
in  eq. \ref{eq:c0d}. However we will also consider the plaquette
correlators eq. \ref{eq:cc}, which will require leaving the vacuum
contribution as a fitting parameter.

The latter - fits to the plaquette correlators -
 will be our only analysis in the case of the critical propagators, since
they  show  a
peculiar behaviour, which we will describe a bit more fully at the end of
this Section.

\begin{figure}
\pic{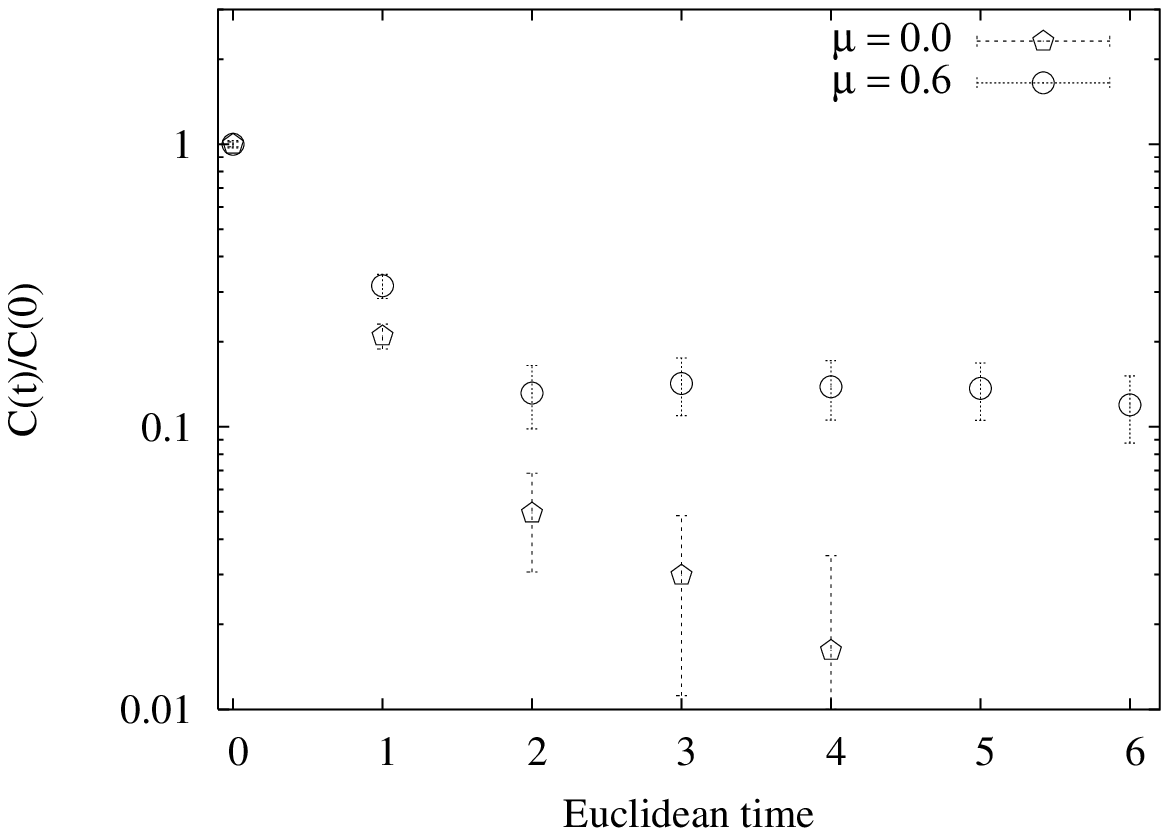}

\pic{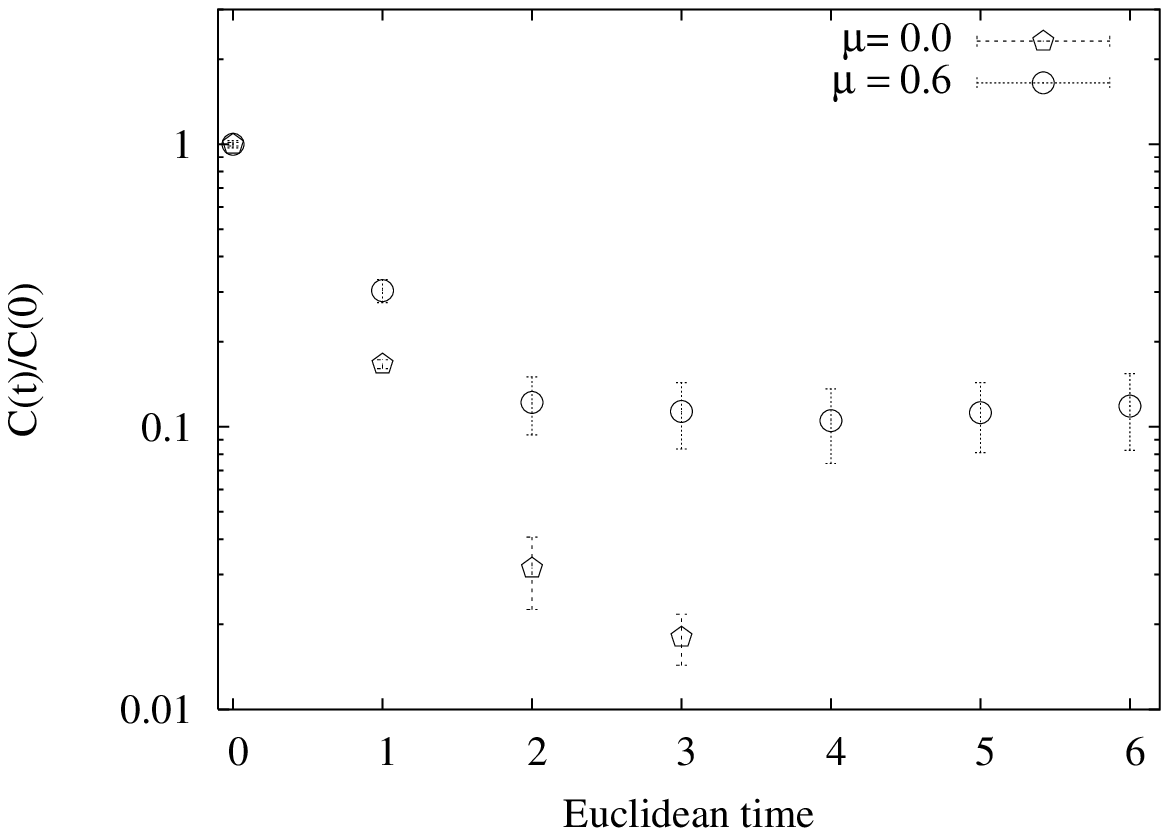}
\caption{Propagators of the scalar glueball at mass $m_q = 0.05(0.07)$, 
upper (lower)
in the normal and superfluid phase, for smearing parameters $w=0.3, T_s =1$,
normalised to one at zero distance. 
Either diagrams demonstrate the decrease of the mass of the 
lowest excitation in the scalar glueball channel when going 
from the normal to the superfluid phase.}
\label{fig:confronta_masse}
\end{figure}

To get a first feeling about the results
in the two phases  we plot in Fig. \ref{fig:confronta_masse} the 
$0^{++}$ glueball correlators (eq.\ref{eq:c0d}), for the two masses we have
inspected, smearing parameters as indicated, and for the two phases.
Other smeared data behave similarly.  
We normalise the propagators to one at zero distance
(the behaviour of the amplitudes will be discussed below) to
put in evidence the decay rate as a function of euclidean
time. It is immediate to notice that the glueball correlator
decays much slower in the superfluid phase, indicating that
the lowest excitation in the gluonic channel becomes
lighter.

In the remaining part of this section 
we discuss first the behaviour of the amplitudes of the correlators,
then we analyze  their long
Euclidean time behaviour. 
The former indicate the critical point,
the latter allow a quantitative estimate of the spectrum in the
two phases. 

\subsection{Amplitude}

\begin{figure}
\pic{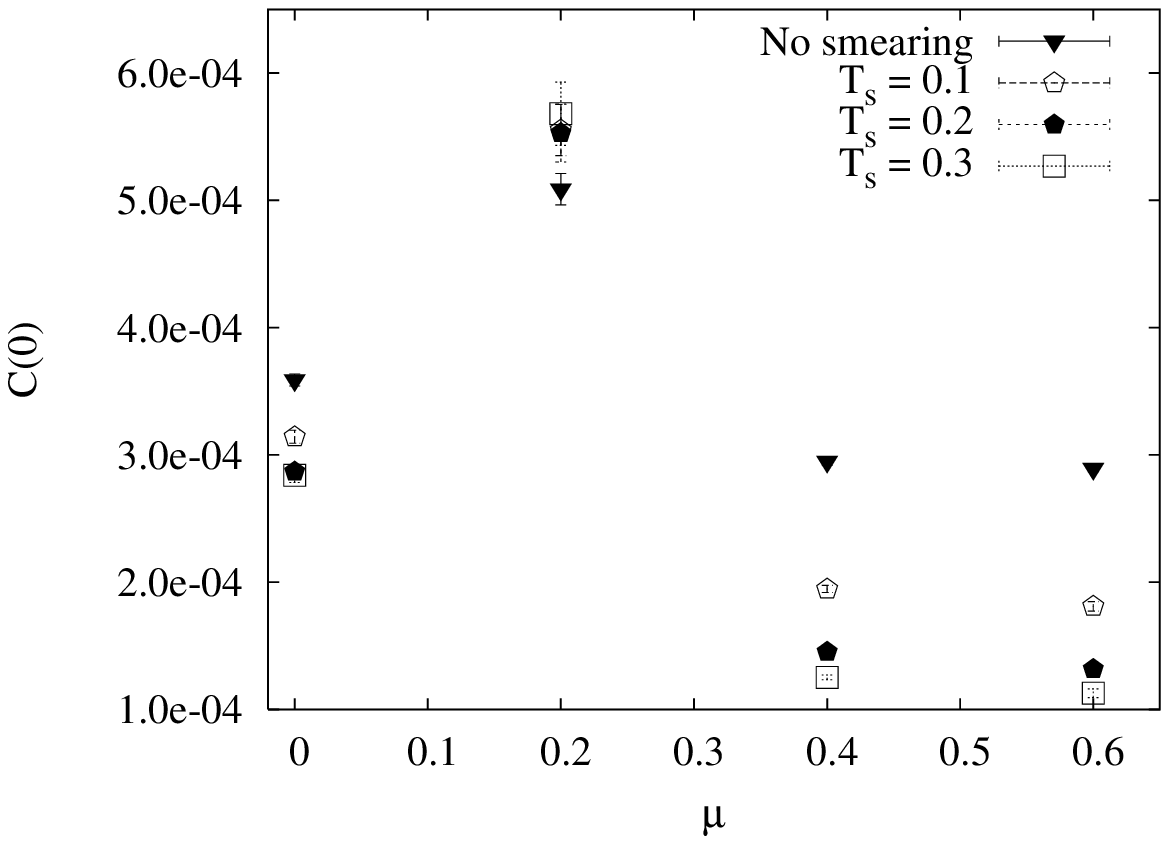}

\pic{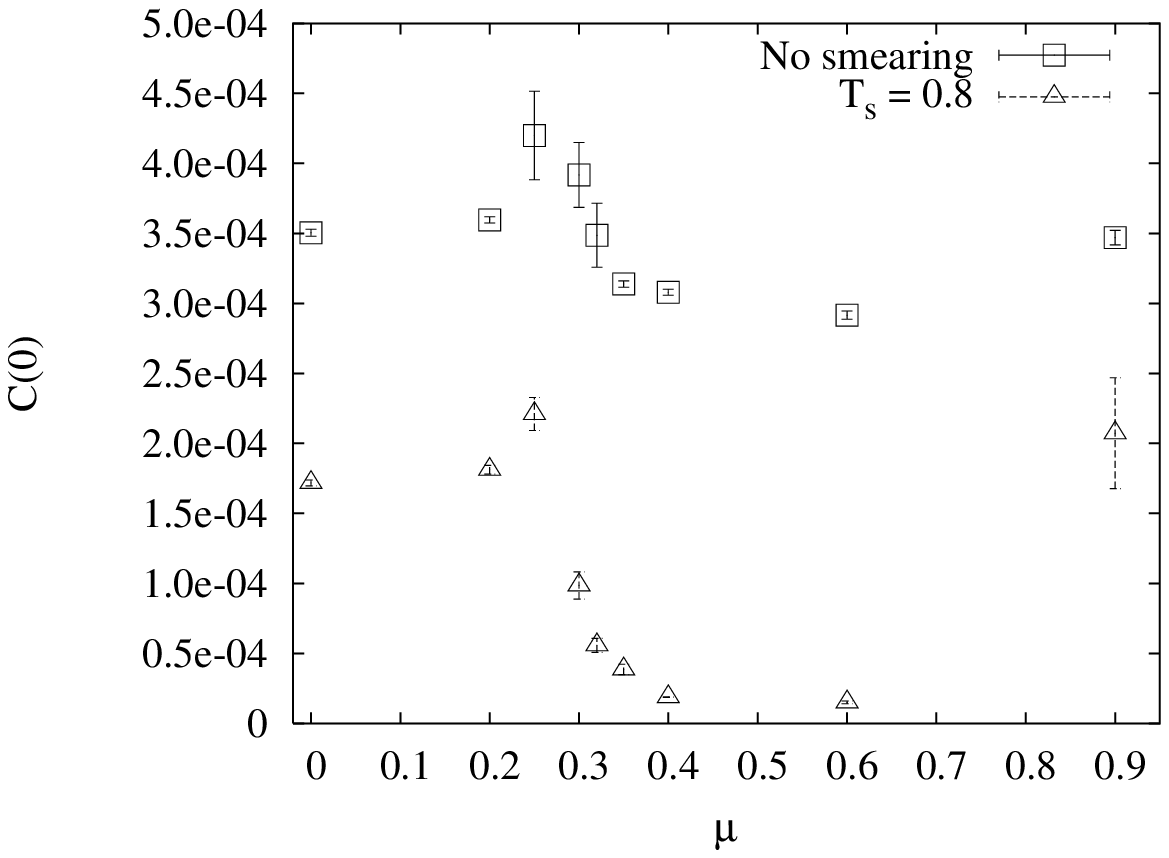}
\caption{Amplitudes of the correlator as a 
function of $\mu$ for raw and smeared data, and quark mass $m_q= 0.05(0.07)$, upper (lower) diagram. The amplitudes, related with the plaquette susceptibilities, peak at the critical point 
$\mu_c = m_\pi/2$ and decrease in the superfluid phase. 
These features demonstrate
that the phase transition is visible in the gluonic sector, and that 
a high density of baryons affect the gluon mass spectrum in the superfluid 
phase.}
\label{fig:amplimu_m5}
\end{figure}

Fig.~\ref{fig:amplimu_m5} shows the amplitudes  $C(0)$
of the correlators, eq.~\ref{eq:c0d},  as a function 
of the chemical potential, for the raw data, and for a representative
value of $T_s$. Other smeared data behave similarly. 
Remember that our estimates of the pion mass, as well as previous
results~\cite{Hands:1999md}, indicate $\mu_c (m_q = 0.05) = 0.28(1)$ 
and $\mu_c(m_q = 0.07) = 0.32(2)$. We note that 
the amplitudes  peak in correspondence with these values.

This is 
consistent with their definition. In fact, $C(0)$ is
defined as 
\begin{equation}
 C(0) = \frac{1}{N_t} \sum_{\tau = 0}^{N_t - 1}
 ( <\phi ^{0^{++}} (\tau) ^2 > - 
<\phi ^{0^{++}} (\tau)>^2) 
\end{equation}
hence, it is 
related with the plaquette susceptibility, which is a standard indicator
of criticality.

A second observation concerns the amplitude
in the superfluid phase,
where the amplitude itself gets smaller.

Moreover, see again
Fig.~\ref{fig:amplimu_m5}, which shows that the
effect of the smearing becomes more pronounced in the superfluid
phase.
Smearing cancels ultraviolet fluctuations and preserves long distance
physics.  Na{\"\i}vely, any observable which has only short distance
components would disappear after smearing. Since in the high
density phase the residual amplitudes is comparably smaller
than in the low density phase, one might conclude that the relative
weight of long distance, low mass components in the high density
phase is comparatively smaller than in the low density phase -
in other words, if the amplitudes would become zero, the glueball
would decouple completely.

At even larger $\mu$ value the results increase and approach the quenched
ones, in agreement with the conclusions in ref.~\cite{Alles:2006ea}.

The main conclusion from the consideration of the amplitudes is that the
either the transition and the superfluid phase are clearly seen also in a pure
gluonic channel. This demonstrates the effect of a finite
density of baryons on the gauge fields: since our operators do not
contain explicitly the chemical potential, this dependence is only
possible if the gauge field themselves are affected by a density of baryons,
both in a pure phase and at the critical point.

These results indicate a sizable modification
in the gluonic sector induced by a finite density of baryons. It is
then very natural to ask ourselves if the confining properties of
the theory are affected as well. To address this question, even at
a very preliminary level, we study the behaviour of the Polyakov loop,
which is an exact order parameter for confinement in the quenched theory,
and an indicator of enhanced screening properties with dynamical fermions.

\begin{figure}
\pic{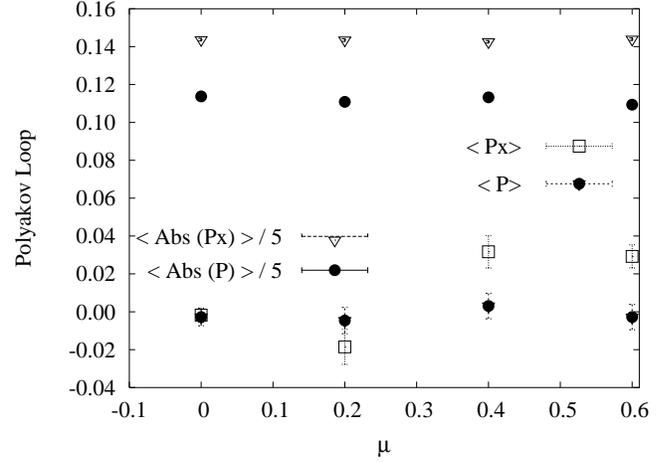}
\caption{Polyakov loop P, and spacial Polyakov loop P$_x$ a
t mass  $m_q= 0.05$ as a function of $\mu$.
It is also shown the (rescaled) average of the local absolute
values of the same observables.}
\label{fig:pol}
\end{figure}

We show in Fig.~\ref{fig:pol} 
the behaviour of the Polyakov Loop at mass $m_q= 0.05$ as a function of
$\mu$. The effect, if any, is very modest and more visible for the spatial
loop. As a comparison, we show also the spatial Polyakov loop.
To attempt to put in evidence bubble or ordered structures we
have also computed, and plotted in the same
figure,  the average of the absolute 
modulus of the local loops. They stay constant
as well. These results suggest that the confining properties of the 
superfluid
 phase appearing at $\mu_q > m_\pi/2$ 
are the same as those of the hadronic phase, in agreement with 
 \cite{Hands:2006ve}. 
In the same paper \cite{Hands:2006ve}
the Polyakov loop is shown to increase from near-zero values at
even larger $\mu$, implying that the confining property of the vacuum is
lost at very high density. In this paper we only consider the
BEC superfluid phase  appearing at $\mu_q > m_\pi/2$, and we do not
discuss any further
transition or crossover at larger $\mu$.

\subsection{Glueballs in the normal phase}
\begin{figure}[h]
\bc
\pic{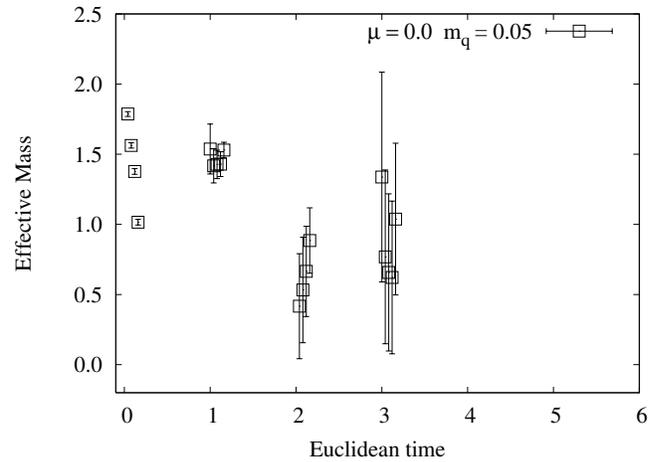}
\caption{Glueball effective masses
as a function of Euclidean time, for $\mu=0.0$  and $m_q = 0.05$.
Results for  $T_s=N_s\times w = 0, 0.1,0.2,0.3,1.2$ are plotted
slightly displaced one from the other, from left to right.}
\label{fig:gmass0500}
\ec
\end{figure}

\begin{figure}
\bc
\pic{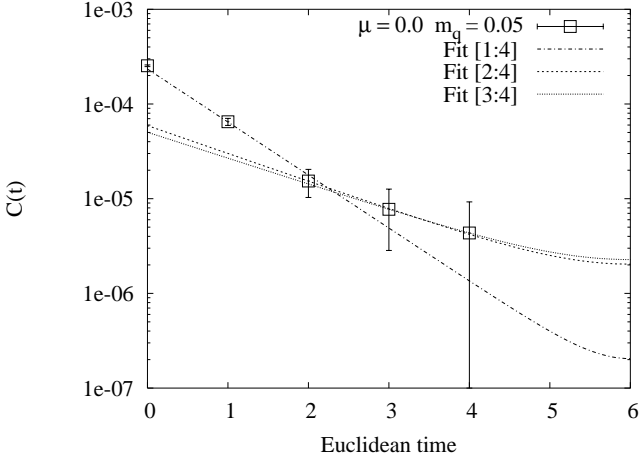}
\caption{Details of the fits 
for different time interval for $\mu=0.0$, $m_q = 0.05$,
smearing weight w = .2 and $N_s$ = 2 ($T_s = 0.2$).
Fits to propagators with other smearing parameters behave similarly,
see text for details, and Table \ref{tab:res0pp} for a summary of the results.}
\label{fig:022_mum5}
\ec
\end{figure}

The measurements of the glueballs in normal conditions are notoriously
difficult, since the glueballs themselves are rather heavy. 
To be more specific, the plaquette correlators are basically constant within
errors. To put in evidence the exponential decay one needs to subtract the
long distance plateau. This is accomplished by  
exploiting the cluster property, and subtracting the square of the plaquette.
Once this is done a signal for the mass - the value
of the exponent - can be detected, while the errors of the fits on the 
amplitudes remain large. This procedure is justifiable only when 
all of the excitations are massive, which we expect to be the case
in the pure phases, while in the critical region the presence of soft,
massless excitations might call for some additional care. In any case,
the validity of the procedure can be assessed a posteriori from the
quality of the exponential fits.  

We 
present some results here for the sake of comparison with the analysis
in the superfluid phase, which is the main goal of this study.

We have started with an effective mass analysis, where a time-dependent
mass $m_{eff}( \bar t)$ is extracted according to eq.~\ref{eq:eff}.

We have studied the smearing dependence of the effective masses 
extracted at different Eulidean times. The effective mass should approach
the true lowest excitation either at large $T_s$  and/or large 
Euclidean time,
and considering the simultaneous dependence on these parameters should
in principle allow a good control on this measure.

We show the results for the effective masses extracted from
correlators at $m_q = 0.05$, for $T_s = 0.0, 0.1, 0.2, 0.3, 1.2$ 
in Fig. \ref{fig:gmass0500}.
The results are very noisy, anyway
it is clear that in the normal phase the glueball is heavy as expected,
and our results allow us to place at least a lower bound for its value.

Next, we performed direct fits of the data to the form of eq.~\ref{eq:eff}.
 In attempt at improving the quality of the results we 
considered, along side with $T_s = 0.6$ and $1.2$, two
higher values of $T_s$.

For $m_q = 0.05$ the fits for $t \ge 2$ were satisfactory, while those
including $t=1$ were not adequate to describe
the largest time distances, as apparent also from a higher $\chi^2$. 
As an example, 
in Fig. \ref{fig:022_mum5} 
we show some fits for the
parameters w=0.2 and $N_s = 2$ ($T_s = 0.4)$. 
The resulting masses where 1.32(2) for the fit performed in the time
interval
[1:4] , which is clearly missing the large time separation behaviour,
while the fits in the interval [2:4], and [3:4] gives 0.68 and 0.63,
in good agreement with each other, but errors on the fit 
can be hardly estimated, given the large errors on the propagators
themselves. The
behaviour demonstrated in Fig. \ref{fig:022_mum5} is  generic for
the $m_q = 0.05$, and for $T_s = .6$ and $1.2$ for $m_q = 0.07$.

For the two larger values of $T_s$ which we have considered, $T_s =
1.6$ and $T_s = 2.0$   the fit of propagators with a larger
$T_s$ in the interval [2:5] are of a poor quality, while the fits 
in the interval [1:5] offer a good description of the data, in agreement
with results at smaller $T_s$ , for $t \ge 2$. In most cases the propagators
for Euclidean time 5 and 6 where zero within errors, as expected given the
large values of the mass, and including them
does not change the quality of the fits. 

We summarize the results  from the fits at $\mu=0$  in Table~\ref{tab:res0pp}.
For $m_q = 0.05$ , and all $T_s$, we quote  results and errors from the fits
in the [2:4] interval.

For  $m_q = 0.07$, a direct comparison of the results show
that the glueball masses are indeed heavier than those 
at $m_q = 0.05$. This implies that
the propagators at $m_q = 0.07$ can be compatible with zero
already at a distance 4, and when this happens is very difficult to
obtain reliable estimates. For $T_s = 0.6, 1.2$
we quote the results obtained by fitting data in the [2,5] interval.
They are consistent with those in the [3,5] interval, which have
however a lower significance. For
 two higher $T_s$ values we quote the results of
the fits in the [1,5] interval (the results in the [2,5] interval
are $m = 2.03(50)$ , and $m = 2.24(65)$, compatible in the large
errors), having in mind that these might well
represent  upper bounds rather then asymptotic estimates.

\begin{table}
\caption{Results for the $0^{++}$ Glueball  at $\mu=0$}
\label{tab:res0pp}
\begin{tabular}{|l|l|l|l|l|}
\hline
\hline
\multicolumn{5} {|c|} {$m_q = 0.05$}\\
\hline
& $T_s = .6$ & $T_s = 1.2$ & $T_s = 1.6$ & $T_s = 2.0$ \\
m & 0.85(12) & 0.98(12) & 1.02(12)  & 1.05(12)  \\
$\chi^2$ & 0.08  &  0.08  &  0.08           & 0.08     \\
\hline
\hline
\multicolumn{5} {|c|}{$ m_q = 0.07$} \\
\hline
& $T_s = .6$ & $T_s = 1.2$ & $T_s = 1.6$ & $T_s = 2.0$ \\
m & 1.02(8) & 1.7(4) & 2.05(8)  & 2.08(8)  \\
$\chi^2$  & 1.93  &  0.2  &  0.2           & 0.2     \\
\hline
\end{tabular}
\end{table}

\begin{figure}
\pic{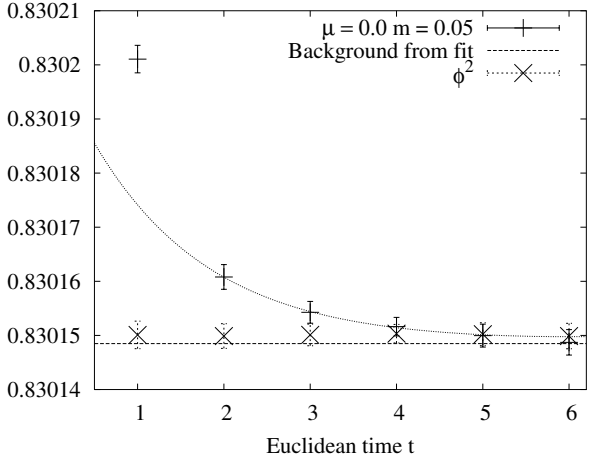}
\caption{Fit of the plaquette correlator, $m_q = 0.05, \mu = 0.0$, 
$T_s = 2.0$.
The mass value 
is consistent with the estimate from the effective mass analysis
and the fits with subtracted background and the fitted vacuum
coincide with errors with the fitted measurement.}
\label{fig:comp}
\end{figure}

Finally, we consider also the plaquette correlators eq. \ref{eq:cc}
where the vacuum contribution was left as a fitting parameter. 
To compare the two fitting forms we considered first the same
fitting interval [2:4], as well as an extended to interval [2:5] ,
and we show the results for $m_q = 0.05$ in Table~\ref{tab:res0ppfit}, while
Fig. \ref{fig:comp} illustrates the generic features of the results.
All in all, subtracting the background allows to constrain one parameter
fit to zero, thus achieving a more precise fit, but the results
within the larger errors are in substantial agreement. In particular,
the fitted vacuum coincides with its direct calculation.

\begin{table}
\caption{Results for the $0^{++}$ Glueball  at $\mu=0$ from
fits with an open vacuum contribution. }
\label{tab:res0ppfit}
\begin{tabular}{|l|l|l|l|l|}
\hline
\hline
\multicolumn{5} {|c|} {Time interval [2:4]}\\
\hline
& $T_s = .6$ & $T_s = 1.2$ & $T_s = 1.6$ & $T_s = 2.0$ \\
m & 0.93 & 0.82 & 0.84  & 0.89  \\
\hline
\hline
\multicolumn{5} {|c|}{Time interval [2:5]} \\
\hline
& $T_s = .6$ & $T_s = 1.2$ & $T_s = 1.6$ & $T_s = 2.0$ \\
m & 0.4(4) & 0.62(25) & 0.68(12)  & 0.73(11) \\
\hline
\end{tabular}
\end{table}

\subsection{Glueballs in the superfluid phase}

\begin{figure}
\pic{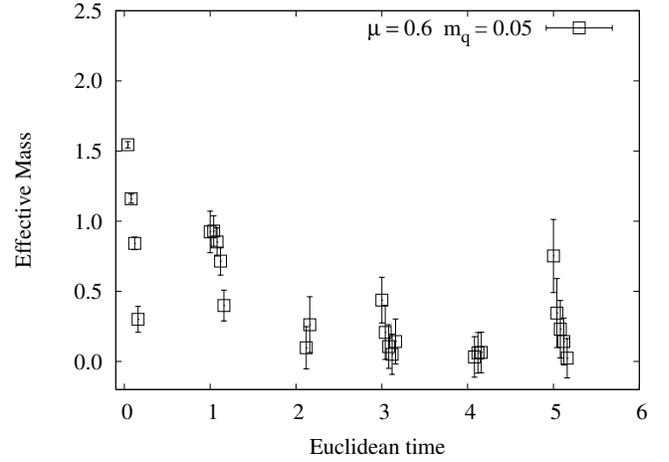}
\caption{Glueball effective masses
as a function of Euclidean time, for $\mu=0.6$  and $m_q=0.05$.
Results for  $T_s=N_s\times w = 0, 0.1,0.2,0.3,1.2$ are plotted
slightly displaced one from the other, for increasing values
of  $T_s$.}
\label{fig:gmass0506}
\end{figure}

We have first performed an effective mass analysis. We show in
Fig.~\ref{fig:gmass0506} a subset of the results, for quark mass $m = 0.05$,
and $\mu = 0.6$, and selected values of $T_s$. $\mu = 0.06$ is well below
the saturation threshold, the baryonic density is less than one (see
e.g. \cite{Lombardo:1999gr}) and the results for the amplitudes are far 
from the quenched ones,
which are approached only when $\mu$ approaches 0.9, as discussed above.

At a variance with the behaviour observed at $\mu=0.0$,
here a plateaux begins to emerge, indicating that by increasing
the time distance, and/or the number of smearing steps we can approach the
asimptotic behaviour representing the true value of the mass. 

\begin{figure}
\pic{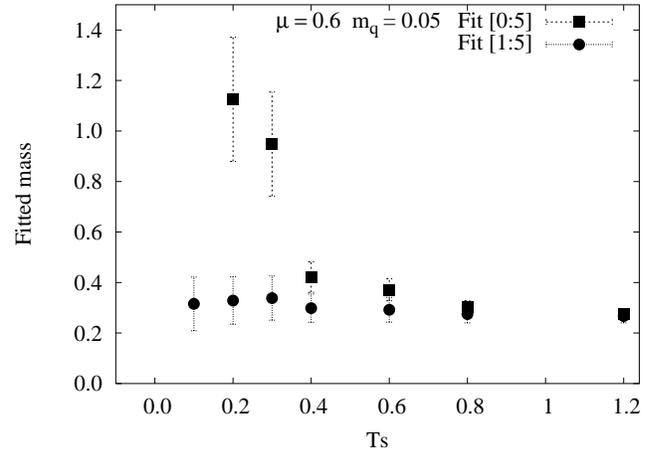}
\caption{Fit results as a function of the smearing parameter, for the
two interval considered. The results confirm the ones seen in the
effective masses, with smaller errors }
\label{fig:fitmu6m5_2}
\end{figure}

The fits proven to be rather stable, and indeed we observed convergence 
over a rather large interval of $T_s$, the convergence to the final results
being faster in the time interval $[1:5]$. We underscore that for $t = 5$
the signal is very weak, so the interval $[1:5]$ is the smallest
which can be fitted in a meaningful way. We summarise the results for the
masses over a large set of $T_s$ in Fig.~\ref{fig:fitmu6m5_2}, which
demonstrates the nice convergence of the results.  The effective mass analysis
give also compatible results, within the larger errors.

Finally, 
Table~\ref{tab:glubmunot0} collects the results for the glueball masses obtained for the
fits with $\mu \ne 0$.

\begin{table}[t]
\caption{Glueball masses at $\mu \ne 0$
extracted from different fits }
\label{tab:glubmunot0}
\begin{tabular}{|c|c|c|c|c|}
\hline
\multicolumn{5} {|c|} {m= 0.05}\\
\hline
& $T_s = .6$ & $T_s = 1.2$ & $T_s = 1.6$ & $T_s = 2.0$ \\
$\mu = .2$ &  - &  -           &  -           & -     \\   
$\mu = .4$ & -  & 1.34 (6)     & 1.16(4)    & 1.07(4)      \\
$\mu = .6$ & 0.31(8)  & 0.27(3)   & 0.26(2)   & 0.27(2)      \\
\hline
\multicolumn{5} {|c|} {Reduced $\chi ^2$  associated to the fits above} \\
\hline
& $T_s = .6$ & $T_s = 1.2$ & $T_s = 1.6$ & $T_s = 2.0$ \\
$\mu = .2$ & -  &  -           &   -          &  -    \\   
$\mu = .4$ & -  &   0.6          &   0.37          &   0.30   \\
$\mu = .6$ & 1.2  & 0.22             &  0.16       & 0.14     \\
\hline
\hline
\multicolumn{5} {|c|} {m= 0.07}\\
\hline
& $T_s = .6$ & $T_s = 1.2$ & $T_s = 1.6$ & $T_s = 2.0$ \\
$\mu = .2$ &2.02(38) & 2.05(28)& 2.06(24)  & 2.05(27)     \\   
$\mu = .4$ & -  & 1.29 (34)   & 1.40(4)    & 1.15(7)      \\
$\mu = .6$ & 0.41(8)  & 0.35(4)   & 0.35(3)   & 0.35(3)      \\
\hline
\multicolumn{5} {|c|} {Reduced $\chi ^2$  associated to the fits above} \\
\hline
& $T_s = .6$ & $T_s = 1.2$ & $T_s = 1.6$ & $T_s = 2.0$ \\
$\mu = .2$ & 1.73 &  2.77           &   2.13   &  1.73    \\   
$\mu = .4$ & -  &   1.18          &   0.03          &   0.17   \\
$\mu = .6$ & 2.36  & 1.06             &  0.82       & 0.78     \\
\hline
\hline
\end{tabular}
\end{table}

Also in this case we performed fits with an open vacuum contribution,
choosing the same interval [1:5]  where the results of the standard
fits proven to be stable with increasing smearing parameter $T_s$,
and in good agreement with effective masses estimates.

\begin{figure} 
\pic{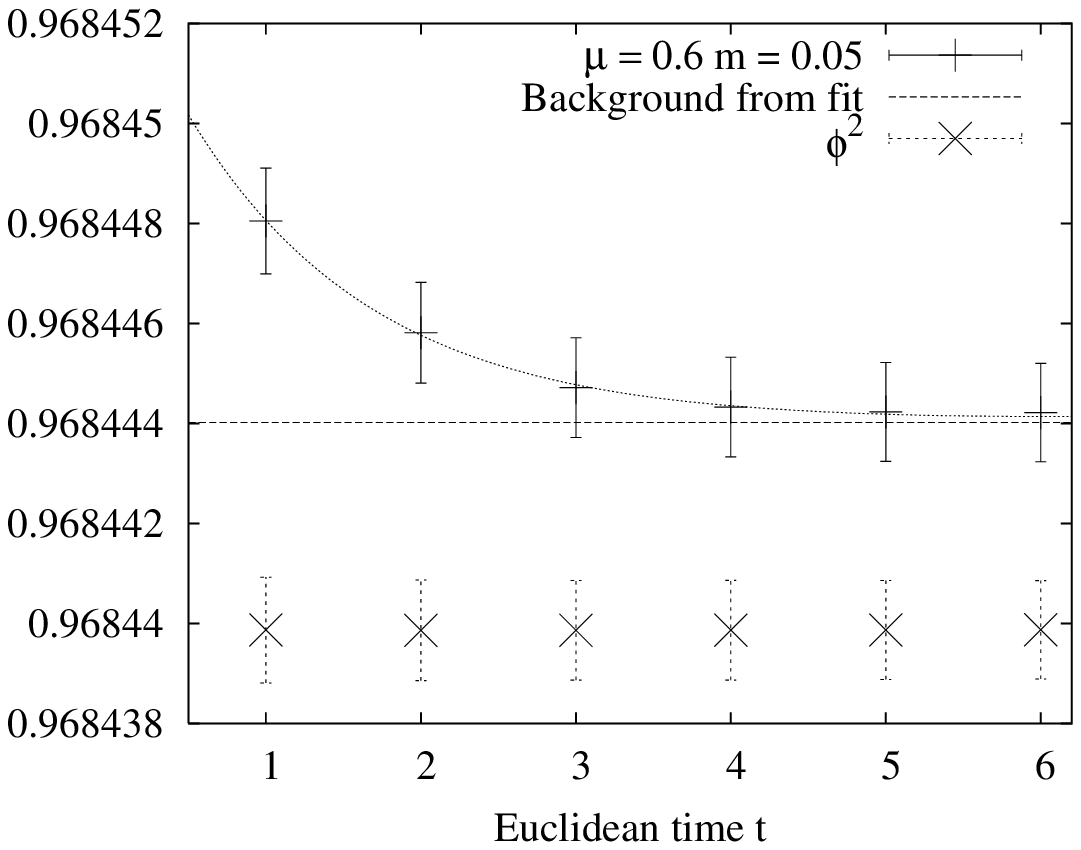}
\pic{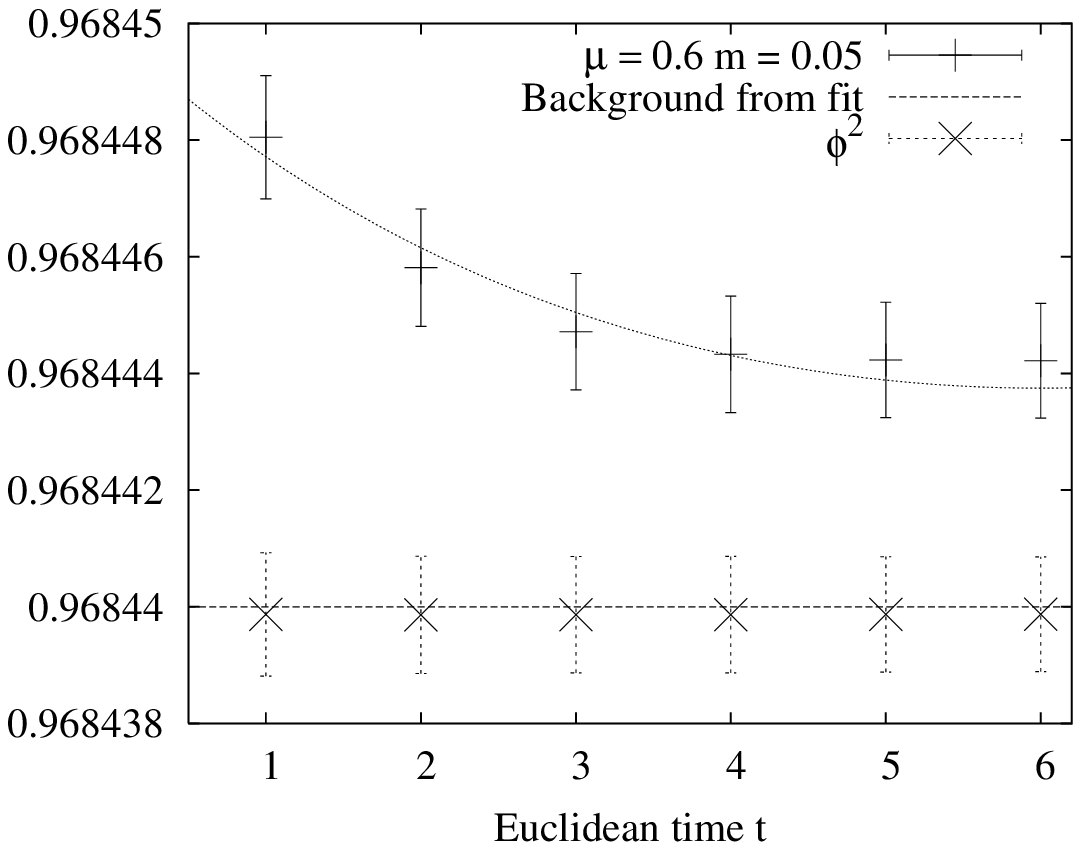}
\caption{The plaquette correlator at $m_q = 0.05$, 
 $\mu = 0.6$, $T_s = 2.0$ with
superimposed the fit with an open background.  The fitted
value of the vacuum contribution and its direct measurement
are shown as well, demonstrating a mismatch and an overall
inconsistence of this fit (top). The same, but with the
mass value constrained to that measured in the fit with
a subtracted background. The quality of the fit is poorer,
but this time there are no inconsistencies (bottom)}
\label{fig:mis20mu06}
\end{figure}

The quality of the fits was very good, however the mass results turned out
to be larger that those obtained by fitting the propagator with
the subtracted vacuum contribution. In addition to this, the fitted
vacuum was definitively much larger than the direct measurement.
Since the fitted mass was rather large, we should expect not only
the agreement between fitted vacuum and direct estimate, but also
an asymptotic behaviour of the propagator reaching a plateaux
given by such a common value, 
see Fig. \ref{fig:mis20mu06}, top.

 This is not what we are observing:
the plaquette correlator levels off at a value which is much
higher than the direct estimate of the vacuum contribution.
This behaviour can be explained by a contribution from a light
mass which is obscured by the background.

To shed light on these observations we performed further fits
this time  using a constrained value of the mass equal to the
one coming from the fits with the subracted vacuum contribution.
In those cases we obtained accettable fits, with the correct
value of the vacuum contribution, see Fig \ref{fig:mis20mu06}, bottom.

A tentative conclusion is thus that it is difficult do
disentangle the contribution from a low
mass excitation from  an extra contribution to
the background. 
To exhibit the correct behaviour we have to
constrain the background to the measured value. If we do so,
either the effective mass analysis and the standard fits produce
results consistent with each other, over a large set of
smearing parameters and two different time intervals.

Clearly to confirm this picture we should repeat our measurements
on lattices more elongated, and with a larger size. If the picture
outlined here is correct, on a larger lattice
we should be able to measure a two state
signal, and to observe with confidence
an asymptotic behaviour, stable over several smearing steps.

\begin{table}[t]
\caption{Glueball masses at $\mu = 0.6 $
extracted from different fits with an open vacuum contribution}
\label{tab:glubmu06}
\begin{tabular}{|c|c|c|c|c|}
\hline
\multicolumn{5} {|c|} {m= 0.05}\\
\hline
& $T_s = .6$ & $T_s = 1.2$ & $T_s = 1.6$ & $T_s = 2.0$ \\
$\mu  = .6$ & 1.73(1)  & 1.00(5)   & 0.89(5)   & 0.84(6)      \\
\hline
\multicolumn{5} {|c|} {Reduced $\chi ^2$  associated to the fits above} \\
\hline
& $T_s = .6$ & $T_s = 1.2$ & $T_s = 1.6$ & $T_s = 2.0$ \\
$\mu = .6$ & 0.00001 & 0.027             &  0.03       & 0.05     \\
\hline
\hline
\multicolumn{5} {|c|} {m= 0.07}\\
\hline
& $T_s = .6$ & $T_s = 1.2$ & $T_s = 1.6$ & $T_s = 2.0$ \\
$\mu = .6$ & 1.55(10)  & 1.19(7)   & 0.89(6)   & 0.85(6)      \\
\hline
\multicolumn{5} {|c|} {Reduced $\chi ^2$  associated to the fits above} \\
\hline
& $T_s = .6$ & $T_s = 1.2$ & $T_s = 1.6$ & $T_s = 2.0$ \\
$\mu = .6$ & 0.02  & 0.02            &  0.03       & 0.03     \\
\hline
\hline
\end{tabular}
\end{table}

\subsection{The critical region}
\label{cri}

As mentioned at the beginning of this Section, 
the critical propagators show a puzzling feature, namely the square of the
plaquette turns out to be larger than the minimum of the propagator
itself: if we  subract it, as done in the pure phases, 
the propagator would become negative~\cite{lattice2}. 

It is rather natural to intepret this behaviour as the 
manifestation of massless modes, which develop at the phase
transition. 
In the presence of massless excitations there is no reason to
expect a plateau in the propagators, so the negative region
which would emerge subtracting the square of the plaquette 
has no particular significance.

On the other hand, it is clear that a finite lattice cannot accomodate
such massless excitations. We only know that the propagator should
become periodical, and it is reasonable to assume that it should be
continuous with all its derivatives at the origin, since there
is no discrete spectrum dominating the short distance behaviour.
These considerations suggest to try fits with three parameters $a, b, c$ of the form:

\begin{equation}
C(t) = a \cos ( 2 \pi t / N_t) \cosh [ b ( t - N_t/2)] + c
\end{equation}

\begin{figure} 
\pic{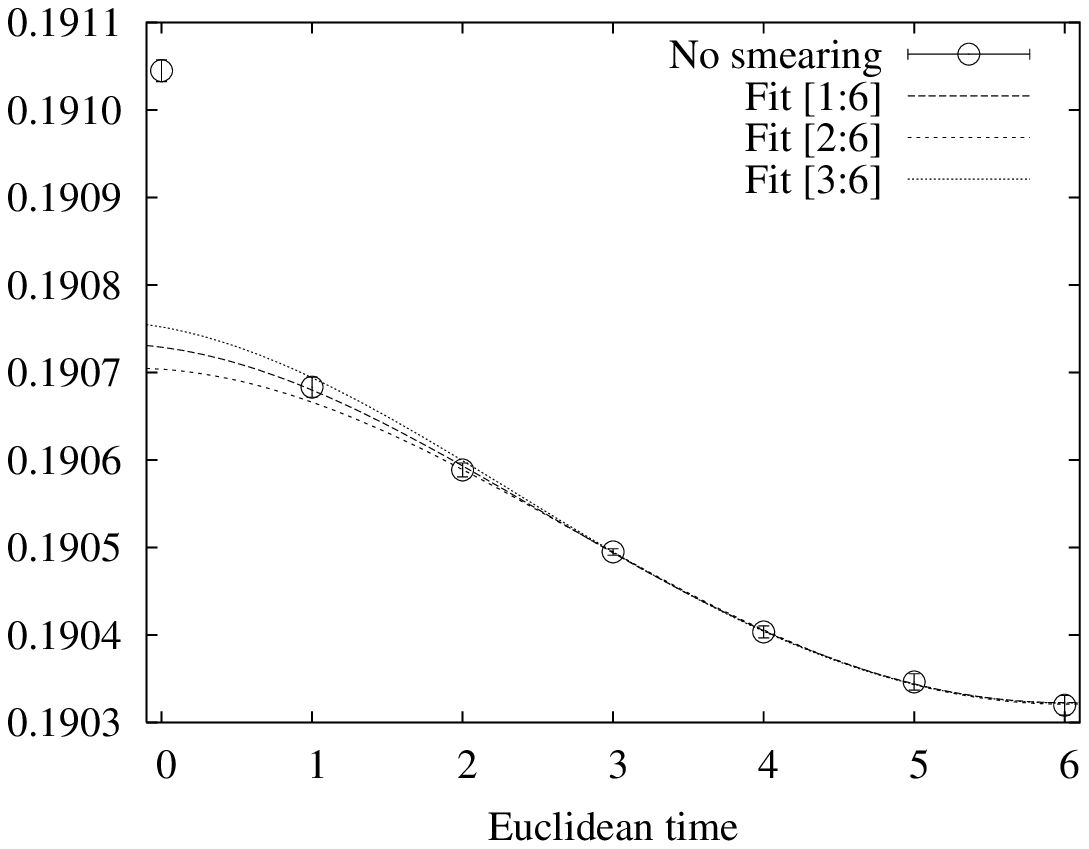}
\pic{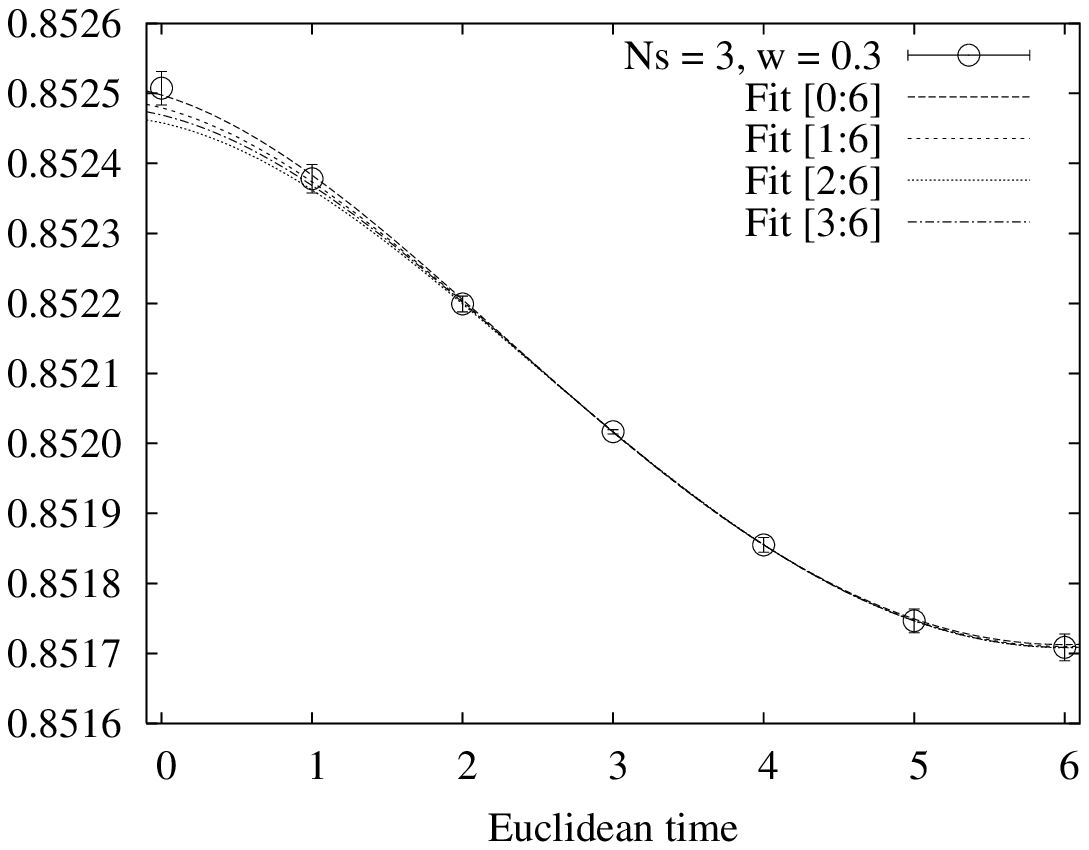}
\caption{The propagator at $m_q = 0.05$, and $\mu = 0.2 \simeq \mu_c$, with
superimposed the fits described in the text. The raw
(smeared) data are shown in the top (bottom) diagram. We obtain
excellent fits to a damped oscillation with period $N_t$, note in particular
the smeared results where a fit constrained to the interval [3:6] 
successfully describes the data in the entire interval}
\label{fig:osc}
\end{figure}

We show in Fig.~\ref{fig:osc}  these fits for the raw
data, and for the smeared ones. The stability of the results is excellent,
note in particular that the fit in the restricted interval $[3:6]$
describes perfectly the data at smaller time distance, including zero
for the smeared results, where the ultraviolet fluctuations are suppressed.

In this picture  the glueball propagator in the critical
region has two complex conjugates poles, whose imaginary part
 $2 \pi / N_t$ goes to zero in the thermodynamic limit.
It is clear that only a  combined  
finite size and finite $N_t$ scaling analysis both 
in the spacial and in the temporal 
direction  might confirm or disprove this intepretation.

\section{Summary and discussions}\label{sec:summary}

We have studied the gluodynamics of 
two color QCD at very low temperature, and nonzero chemical potential. 

We have noted that the Polyakov loop is nearly insensitive to the 
transition from the hadronic to the superfluid phase, in agreement with 
the observations of \cite{Hands:2006ve}, and suggesting that 
the superfluid phase shares the same confining properties as the hadronic
phase in a rather large interval of chemical potential. We have not
addressed the possibility of a deconfining transition taking place
at an even  larger $\mu$ value and the possibility of a BEC/BCS crossover
discussed in \cite{Hands:2006ve}.

We have observed a gluonic transition - signaled by a peak of the 
plaquette susceptibility -  coincident with the superfluid transition
at $\mu_c = m_\pi/2$. Beyond this transition, in the superfluid phase,
the susceptibility  reaches a lower value than in the normal phase. 

We have studied the scalar glueball propagator in the normal and in the 
superfluid phase. In both phases the propagators are amenable to standard
hyperbolic cosine fits, and related effective mass analysis.
In the normal phase the results from two different
fits - one with a subtracted vacuum contribution, and the other with
the vacuum contribution left as a free parameter - give results in
good agreement one with another. In the superfluid phase the subtraction
of the background seems mandatory to obtain coherent results, and in this
case it turns out that the lightest excitation in the scalar glueball channel
is lighter by about a factor two 
in the superfluid phase than in the normal phase. The fit
with a free background would produce a larger mass, but also
an anomalously large fitted
value for the vacuum contribution, not compatible with the direct
calculations. 

At the present stage, we can conclude 
that our results indicate a non trivial modification of the gluonic
vacuum in the low temperature, superfluid phase. 

 To put the above considerations on firmer quantitative
grounds we would need simulations on a larger and more elongated lattice, 
as well as a careful considerations of the mixing effects with 
fermionic channels with the same quantum numbers. 

One interesting open question concerns the
mass dependence of the results. We already know that in the
large chemical potential limit, beyond the saturation
threshold,  we recover the quenched behaviour \cite{Alles:2006ea}.
The quenched limit is also obtained at infinite mass, and in that
case  the gluon fields should not be sensitive to the  chemical
potential.
Then, when increasing the mass, the critical point would move forward and
forward, eventually collapsing with the saturation threshold at infinite
mass, in which case all gluonic observables will remain constant
and equal their $\mu = 0$ value. This suggests the possible existence
of an endpoint of the transition in the chemical potential mass
plane, which will be interesting to explore.

One final comment concerns the interrelation, or lack thereof , of these 
studies with real QCD. 
 Following a large $N_c$ analysis it was proposed \cite{McLerran:2007qj}
that 
the matter immediately above $\mu_c$ is ``quarkyonic'' -  the
authors choses this name to indicate a phase with an unusual realization
of chiral symmetry, which still remains confining.  This phase,
according to the authors, should persist till relatively large temperatures,
and should not break color symmetry, 
at variance with  the ordinary superconducting or CFL phases which
are indeed features of the very low temperatures, and which break 
color symmetry. The two scenarios of
the large $N_c$ a quarkonic phase and the small $N_c$ superfluid phase
share a few common aspects: they are confining, with unusal realization
of chiral symmetry, do not break color symmetry and should persist
till relatively large temperature. The physical phase diagram of real 
QCD at low temperatures should in some sense interpolate between the
two.  It is tempting  to regard the superfluid phase  
as a limiting case of the quarkyonic phase of large $N_c$ QCD. Should this be
true, two color QCD might be a  better guidance to the understaning
of the QCD phase diagram in the physical case than previously thought.

\section*{Acknowledgments}

We wish to thank A.~E.~Dorokhov, S.~J.~Hands, 
E.~M.~Ilgenfritz, M.~M\"uller-Preussker, D.~T.~Son and M.~Testa
for helpful conversations. MPL thanks the 
D.o.E Institute for Nuclear Theory at the University of Washington 
for its hospitality during the completion
of this work. \\
The simulations were performed on the ATLAS computer farm 
in Rome and we thank A. De Salvo for his invaluable help during the development
of this high statistics project.

\end{document}